# Top-Down Mass Spectrometry Imaging of Intact Proteins by LAESI FT-ICR MS

András Kiss[1]; Donald F. Smith[1]; Brent R. Reschke[2]; Matthew J. Powell[2]; Ron M. A. Heeren[1,*]

[1] FOM Institute AMOLF, Science Park 104, 1098 XG Amsterdam, The Netherlands

[2] Protea Biosciences, Inc., 995 Hartman Run Road, Morgantown, WV, USA

*Author to whom correspondence should be addressed. Email: heeren@amolf.nl

**List of abbreviations**

| | |
|---|---|
| AGC | Automatic Gain Control |
| DESI | Desorption Electrospray Ionization |
| ECD | Electron Capture Dissociation |
| ETD | Electron Transfer Dissociation |
| IRMPD | Infrared Multiphoton Dissociation |
| LAESI | Laser Ablation Electrospray Ionization |
| MALDESI | Matrix Assisted Laser Desorption Electrospray Ionization |
| MSI | Mass Spectrometry Imaging |
| NCE | Normalized Collision Energy |
| SIMS | Secondary Ion Mass Spectrometry[1] |

---

[1] Ron M.A. Heeren is a member of the Scientific Advisory Board of Protea Biosciences

**Abstract**

Laser Ablation Electrospray Ionization is a recent development in mass spectrometry imaging. It has been shown that lipids and small metabolites can be imaged in various samples such as plant material, tissue sections or bacterial colonies without anysample pre-treatment. Further, laser ablation electrospray ionization has been shown to produce multiply charged protein ions from liquids or solid surfaces. This presents a means to address one of the biggest challenges in mass spectrometry imaging; the identification of proteins directly from biological tissue surfaces. Such identification is hindered by the lack of multiply charged proteins in common MALDI ion sources and the difficulty of performing tandem MS on such large, singly charged ions. We present here top-down identification of intact proteins from tissue with a LAESI ion source combined with a hybrid ion-trap FT-ICR mass spectrometer. The performance of the system was first tested with a standard protein with ECD and IRMPD fragmentation to prove the viability of LAESI FT-ICR for top-down proteomics. Finally, the imaging of a tissue section was performed, where a number of intact proteins were measured and the hemoglobin α chain was identified directly from tissue using collision-induced dissociation and infrared multiphoton dissociation fragmentation.

## Introduction

The importance of mass spectrometry (MS) based proteomics in the field of biological research has grown constantly over the past two decades and has become a powerful tool for biological analysis. The two main approaches in the field of proteomics are bottom-up and top-down proteomics. Bottom-up proteomics uses different proteolytic enzymes, such as trypsin, to digest the intact proteins into smaller peptide fragments. These peptides are then typically separated and identified by the combination of liquid chromatography and mass spectrometry. Despite its widespread successful application, the method has several drawbacks. First, it is challenging to retain labile post translational modifications (PTM) and to identify different proteoforms[1]. Secondly, in bottom-up proteomics, the sequence coverage of a protein is limited due to the poor fragmentation of some of the peptides and the discrimination of the proteases for certain amino acid residues.

Top-down proteomics[2], however, analyzes intact proteins without any prior protease treatment. Thus, labile PTMs are retained during mass spectrometric analysis. However, multiply charged ions are necessary for most mass spectrometers to enable detection and effective fragmentation. In the overwhelming majority of experiments, this is typically achieved by electrospray ionization (ESI)[3-5]. A mass spectrometer with high mass resolving power is necessary to resolve the isotopic envelopes of the high charge states of the precursor and fragment ions produced in a top-down proteomics experiment. This is required for the proper deconvolution of the complex spectra produced in top-down proteomics. This means that typically Fourier Transform mass spectrometers, such as Fourier Transform Ion Cyclotron Resonance (FT-ICR)[6] and orbital trapping[7, 8] (i.e. the Thermo Fisher Orbitrap) mass spectrometers, are used for top-down proteomics research. These types of instruments combine exceptional mass resolving power with several fragmentation methods, such as collision-induced dissociation (CID), electron capture dissociation (ECD), electron transfer dissociation (ETD) and infrared multiphoton dissociation (IRMPD).

Mass spectrometry imaging (MSI)[9, 10] is a method to simultaneously map the distribution of multiple molecules on complex surfaces. The main advantage of the technique over other imaging techniques is its label free nature. One of the main challenges in the application of MSI for proteomics is the identification of detected protein or peptide ions[11]. The traditional ion sources for mass spectrometry imaging are matrix-assisted laser desorption/ionization (MALDI) and secondary ion mass spectrometry (SIMS). However, these ion sources are not suitable for top-down proteomics measurements because they predominantly produce singly charged ions. Thus, proteins are traditionally identified by the bottom-up approach in mass spectrometry imaging experiments. A recent work by Schey *et al.* combines top-down protein identification and mass spectrometry imaging [12]. In this work, after the imaging of the tissue section by MALDI time-of-flight MS, proteins were isolated by microextraction from certain areas of the tissue, which was followed by a traditional top-down MS proteomics workflow. The proteins identified in the top-down MS experiments were subsequently matched to those measured in the MALDI MS imaging experiment, but no identification from the tissue surface was performed.

Recently, ambient pressure ion sources have begun to gain more popularity in the mass spectrometry imaging community. These sources have several advantages over vacuum sources (like MALDI). They allow the analysis of samples that are not vacuum compatible and they simplify sample and source exchange. An additional reason for the elevated interest in ambient ionization sources is their ability to produce multiply charged ions. Most of these ion sources employ electrospray as the main ionization mechanism such as MALDESI[13, 14], DESI[15], nano-DESI[16] and LAESI[17] or have similar ionization mechanisms to ESI, such as Laserspray[18]. The latter has demonstrated to be capable of imaging multiply charged proteins from tissue samples offering promise for top-down proteomic imaging experiments. One example of these is the work done by Inutan *et al*[19], where multiply charged proteins were detected both from standard and tissue with Laserspray ionization and ETD was used for fragmentation. However, they were unable to identify the intact proteins detected from the tissue. While no successful top-down imaging of multiply charged proteins from tissue sections has been published, both nano-DESI and MALDI with special matrices show promise for top-down MS imaging.

Laser Ablation Electrospray Ionization is an ambient pressure ionization method developed in 2007 by the Vertes group[17]. This ionization method employs a mid-infrared laser with the wavelength of 2.94 μm to ablate material from a sample surface. After the initial ablation event, the ablated material interacts with the plume of an electrospray source. This results in the incorporation of the analytes in the charged droplets and the subsequent ionization of the material from the sample surface. Due to the ionization mechanism, multiply charged ions can be produced. The main advantage of LAESI is its matrix free nature. The wavelength of the infrared laser is in the region of the stretching vibrations of the OH groups. Thus, LAESI uses the sample's natural water content as a matrix. LAESI has been used to image or profile several different substrates such as different plant material[20-22], tissue sections[23-25], cell cultures[26], bacterial colonies[27] and textile fabrics[28]. Most of these experiments were done on time-of-flight mass spectrometers. However, the Muddiman group built a LAESI FT-ICR system and demonstrated the systems capability to detect multiply charged proteins such as cytochrome C and myoglobin from both solid and liquid standard samples[29]. In the same work they presented the first example of CID fragmentation of intact protein ions produced with a LAESI ion source from standard samples. The same group later published a modified version of the source for tissue imaging where lipids could be imaged from various tissue sections[30].

Here we present the results of interfacing a commercial LAESI source with an FT-ICR mass spectrometer. For the first time, LAESI is used for imaging of multiply charged proteins directly from biological tissue sections. Subsequent top-down analysis by CID and IRMPD is used for protein identification in the imaging mode. Further, the top-down analysis of proteins from standard liquid surfaces by ECD and IRMPD is presented.

## Materials and methods

### Samples

An 80 µM solution of Cytochrome C standard was prepared in water was used for IRMPD and ECD top-down analysis with the LAESI source. For every measurement, 10 µl of the solution was spotted on a 96 well plate and was measured directly from the surface of the liquid droplets. For tissue imaging experiments, mouse lung (female 9 CFW-1 mouse, Harlan Laboratories, Boxmeer, The Netherlands) was sectioned to 50 µm thick sections in a Microm HM525 cryomicromtome (Thermo Fisher Scientific, Walldorf, Germany) and was deposited on standard microscope slides (Thermo Fisher Scientific, Braunschweig, Germany). The tissue sections were stored at -20 °C until further use and were measured frozen and without any additional sample preparation. A 1:1 mixture of MeOH and $H_2O$ with 0.1 % acetic acid was used as the electrospray solvent in all measurements.

### Mass spectrometry:

Measurements were done on an LTQ-FT hybrid mass spectrometer (Thermo Fisher Scientific, Bremen, Germany) equipped with the IRMPD and ECD option. The LAESI DP-1000 (Protea Biosciences, Inc, Morgantown, WV) ion source was used for all LAESI measurements. A flow rate of 1.5 µl/min and ESI voltage of 4200 V was used for all experiments. The sample was positioned 11 mm below the inlet capillary of the mass spectrometer and 50 mm from the lens of the infrared laser (z-direction). The distance between the ESI needle and the inlet capillary was set to 10 mm.

For the imaging experiments the stage step size was set to 300 µm. At every pixel the ions from 5 laser shots were collected at a laser repetition rate of 10 Hz. The mass spectrometer has been run with the automatic gain control (AGC) turned off. The injection time was set to 900 ms and 1 microscan was collected at every position. The mass resolution was set to 200 000 at *m/z* 400. The tissue imaging experiments have been done in SIM mode, with the mass range set between *m/z* 500 and 1100. This is the mass range where most of the proteins and protein fragments are expected. Additionally, with these settings most of the chemical background ions produced by the ESI source are not injected into the FT-ICR cell which has a beneficial effect on both the spectral quality and the sensitivity of the instrument since the ion-trap and FT-ICR cell are not overfilled with low mass ions. The programmable trigger from the LTQ-FT was used at the start of the analytical scan to synchronize the mass spectrometer, laser firing and X-Y stage movement. For the MS/MS imaging experiments an isolation window of 10 Da was used and the precursor ion was isolated in the ion trap. The CID fragmentation was performed in the ion trap as well with the normalized collision energy (NCE) set to 20. The fragments were detected in the FT-ICR. The IRMPD spectra were measured with the energy set to 20 and the duration to 100 ms. The ECD experiments were done with the energy at 5, the delay set to 30 ms and the duration of 20 ms.

The mass spectrometry data was collected with the Xcalibur software in the Thermo Raw file format. Individual scans were also stored in the MIDAS file format. The spectra were deconvoluted and peaklists were created with the THRASH algorithm[31] built in the MIDAS 3.21 (National High Magnetic Field Laboratory, Tallahassee, FL) data analysis software[32]. ProSight PTM 2.0[33] was used for database search and protein identification. The imaging datasets were converted from the MIDAS raw files to AMOLF developed Datacube format with the Chameleon software package[34] and were analyzed with the Datacube explorer software (FOM Institute AMOLF, Amsterdam, The Netherlands) and in-house developed Matlab code (The MathWorks Inc., Natick, MA).

**Results and discussion**

Cytochrome C standard solution was measured in both direct infusion electrospray mode and directly from the surface of a single liquid droplet with the LAESI ion source to compare the two ionization methods. Supplementary figure S1. shows the comparison of the electrospray and the LAESI spectra. Both measurements provided several different charge states of the protein between 10+ and 19+ charges. These results demonstrate that LAESI is able to provide similar protein spectra as electrospray ionization. However, the LAESI spectrum is shifted to slightly higher charge states. This can be explained by subtle differences in the electrospray conditions between the two sources, or by IR laser induced denaturation.

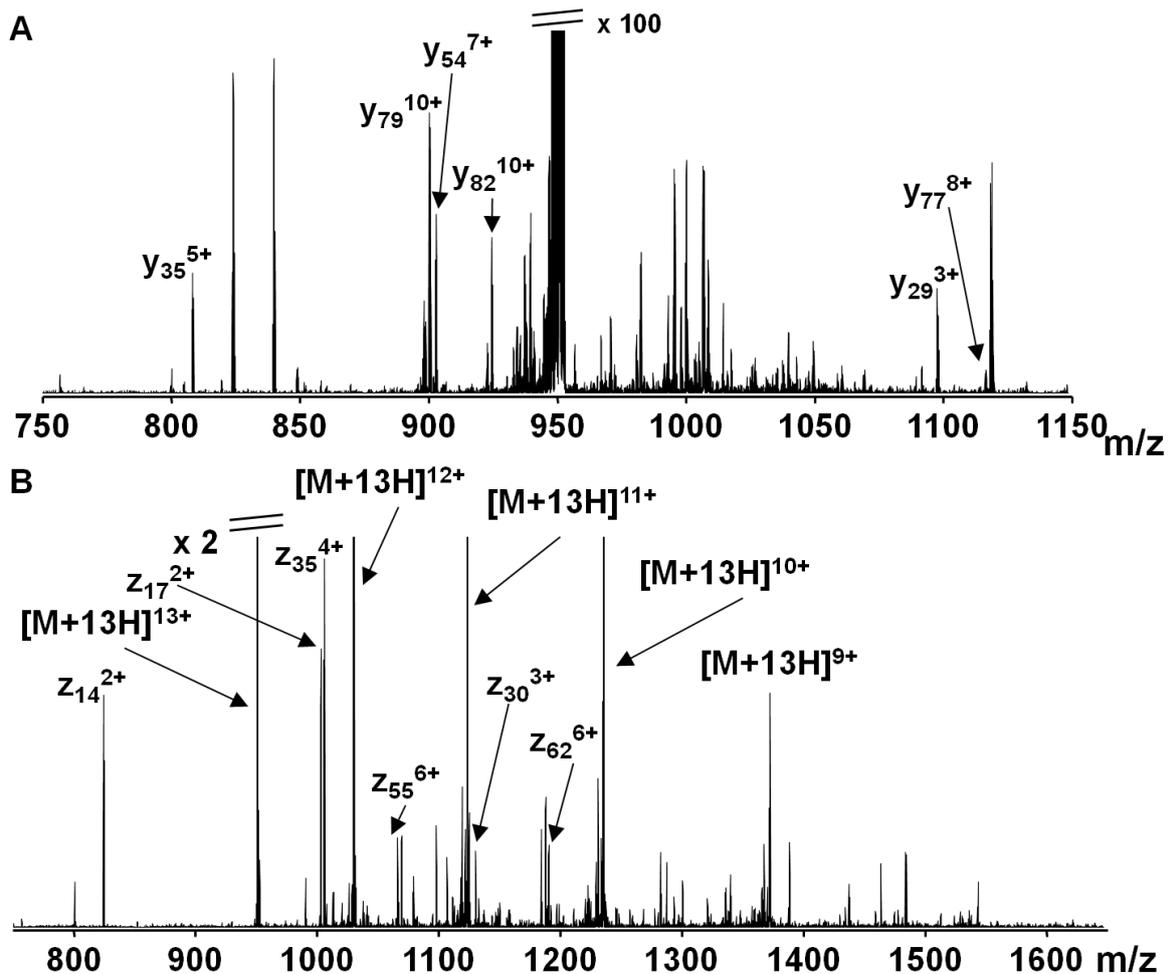

**Figure 1**

The precursor ion of cytochrome C at m/z 951 was fragmented by IRMPD and ECD to prove the suitability of the combination of LAESI and these fragmentation methods for top-down proteomics. As IRMPD and ECD have different fragmentation mechanism they provide complementary information on the protein sequence. IRMPD fragmentation produces b and y protein fragments while ECD fragmentation results in c and z fragments. Also, in ECD fragmentation an extensive charge loss of the precursor ion can be observed. Both the IRMPD and the ECD spectra are shown on Figure 1. The database search after deconvolution of the fragment spectra resulted in the identification of cytochrome C in both cases. As it is shown on Fig. 1, several of the y fragment ions in various charge states were annotated in the IRMPD spectrum and z fragments in the ECD spectrum. The results shown in Fig. 1 confirm that the combination of LAESI with FT-ICR can be used for successful top-down analysis of intact proteins.

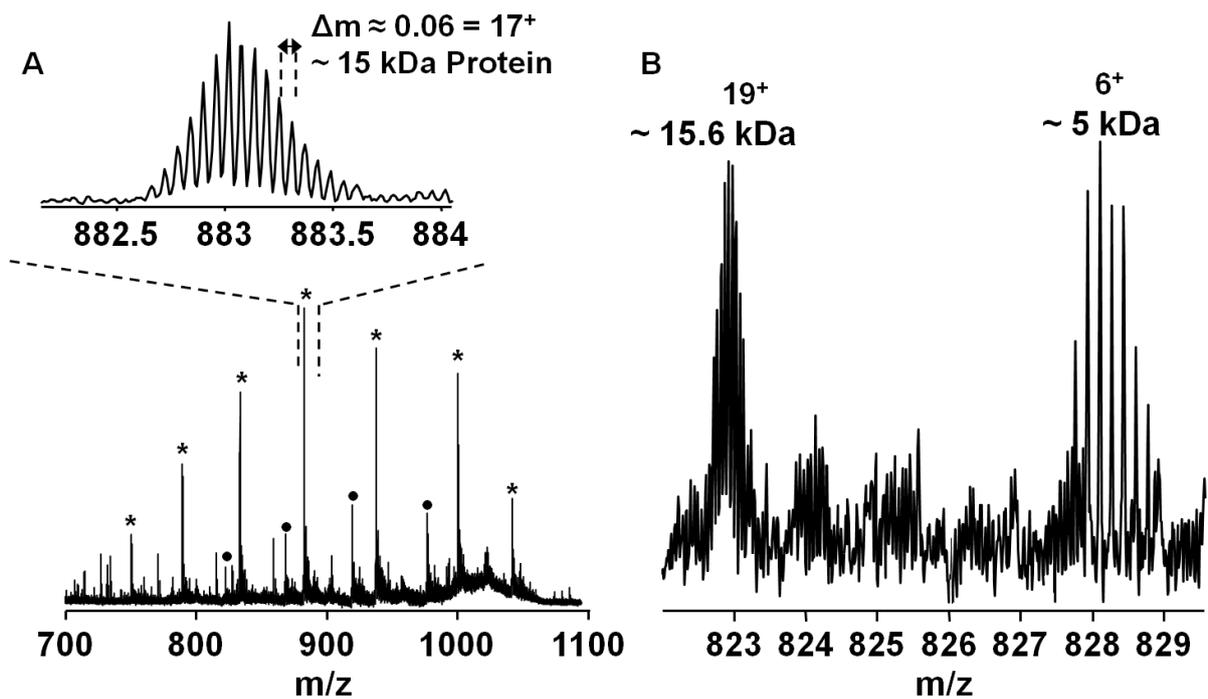

**Figure 2**

Figure 2 shows the summed mass spectrum from an MS imaging experiment on a mouse lung tissue section. The spectrum contains one main charge state series. The mass difference between the isotope peaks of the ion at *m/z* 883 from this charge state series is ~0.06 Da, which means that this ion has 17 charges. Thus the main charge state series is related to a 15 kDa protein. Besides this protein there is also a second charge state series visible which is related to a different protein which has a mass of ~15.6 kDa. Fig. 2b shows the $19^+$ charge state from the lower intensity charge state series and an additional protein detected with 6 charges, which has a mass of ~5 kDa. Fig. 2 demonstrates the viability of LAESI for the analysis of several, unknown intact proteins directly from biological tissue sections. The high mass resolving power of the FT-ICR MS is required to resolve the isotopic distributions and enable proper mass deconvolution.

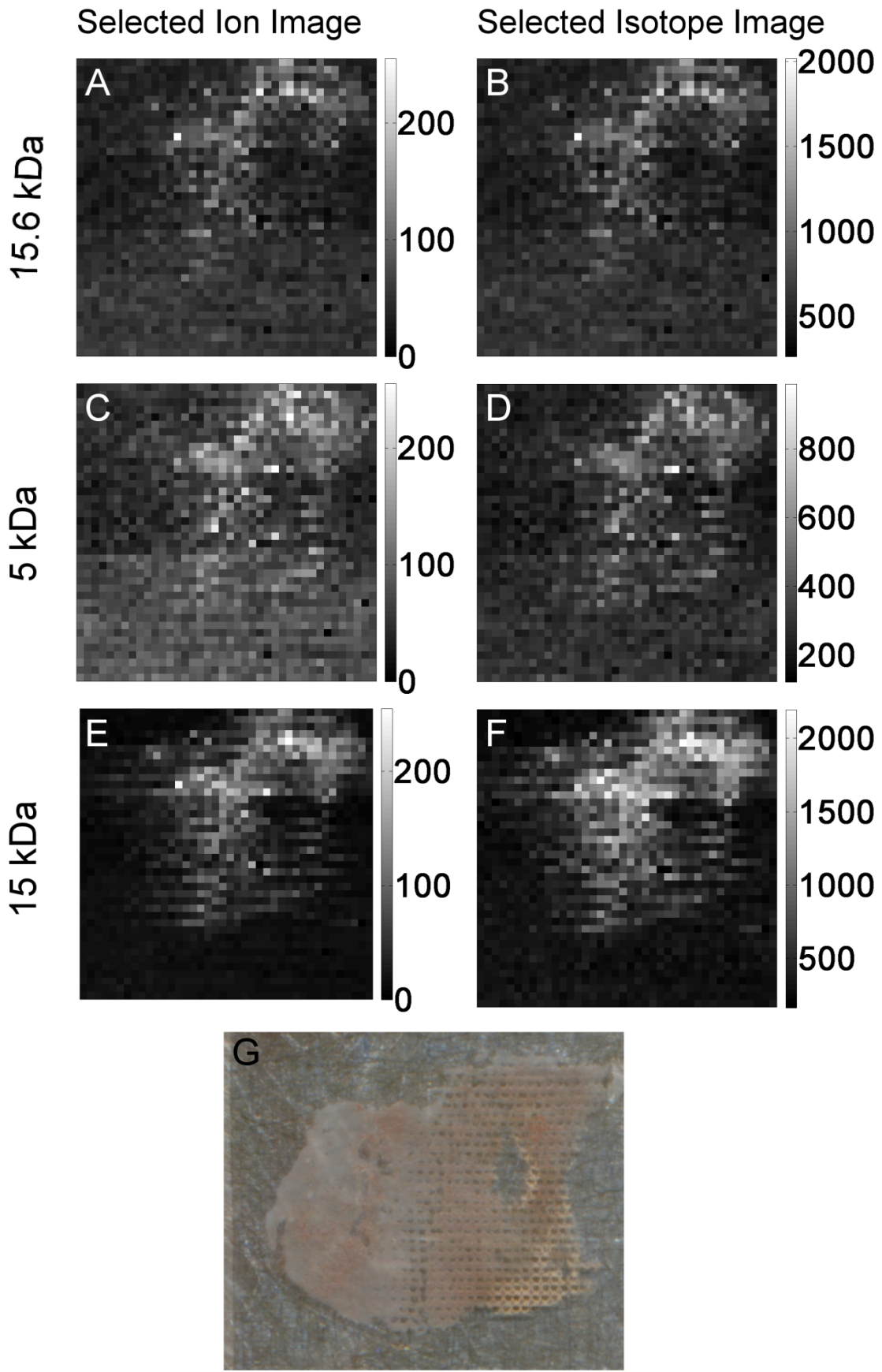

**Figure 3**

Figure 3 shows selected ion images for the ions at *m/z* 823 (15.6 kDa), 827 (5 kDa) and 883 (15 kDa). The images show the distribution of these compounds on the tissue sample, where the hole in the middle of the lung tissue is visible. The compounds are mostly localized in the brown colored areas of the lung section, which means they are likely blood related proteins. Two different approaches were used to plot the distribution of these three ion species. First, the entire isotope distribution was selected for the image. The second approach yields the so called "selected isotope images". This means that the isotope peaks are selected individually and these isotope images are summed together to create the selected isotope image. This second approach is made possible by the high mass resolving power of the FT-ICR mass spectrometer, because it is able to resolve the individual isotope peaks of the highly charged protein ions. As it can be seen on Fig. 3, the selected isotope images provide a better contrast. The selected isotope images additionally minimize the contribution of underlying interferences. Because of the aforementioned advantages the selected isotope images were selected for all the images presented further in this paper.

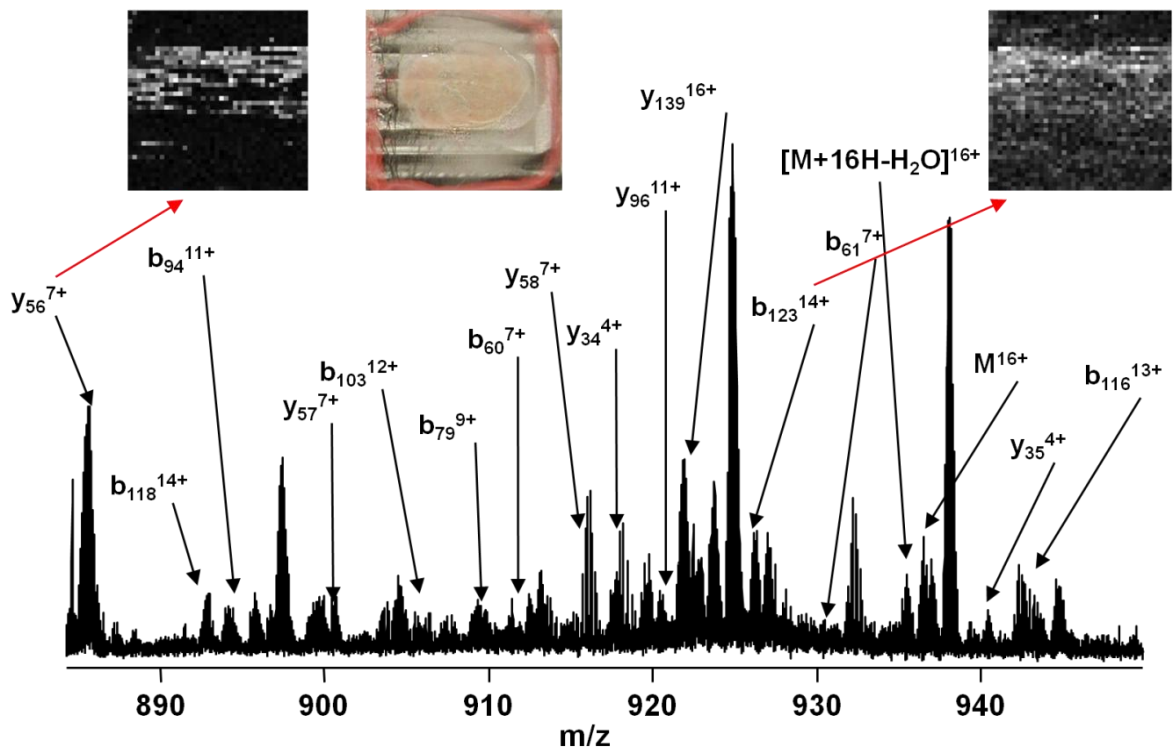

**Figure 4**

The biggest challenge in mass spectrometry imaging of proteins is their identification directly from tissue sections. Thus, the ion at *m/z* 883 has been selected for a CID MS imaging experiment for protein identification. Figure 4 shows the summed mass spectrum between *m/z* 880 and 950 from this CID MS/MS imaging experiment (the broadband mass spectrum is provided in the supplementary material). These MS/MS imaging experiments have two main advantages over profiling MS/MS experiments. First, all fragment ions have an image and secondly the larger number of MS/MS scans result in better statistics and thus better mass spectra. Therefore, the signal-to-noise of the fragment ions is better in the imaging experiments. The spectrum proved to be very information rich. After deconvolution and a subsequent database search in Prosight PTM 2.0, the protein was identified as hemoglobin α with a p value of $1.52*10^{-10}$ and with 19 fragments identified in absolute mass search mode. If the peaklist is searched against the acetylated hemoglobin α sequence from the Uniprot database in single protein mode, then the p value improves to $4,22*10^{-30}$ and 52 fragment ions are identified. The acetylation site was identified as the serine at the 68 position. The number of annotated fragment ions was further improved by the comparison of the identified fragments from the CID fragmentation and the list of the identified fragments from an IRMPD imaging experiment discussed in details in the next paragraph. In this way, the fragment ions where the difference between the theoretical and the experimental mass values was ± 1 Da, due to the deconvolution of the multiply charged ion peak, can be manually annotated.

The identity of the protein is in good agreement with the results from the MS imaging experiment which showed that the protein has a higher intensity in the brown colored areas of the sample. This color is mostly related to blood in tissue sections. Also, this identification is in accordance with the biological role of the tissue, where oxygen is transported into the blood stream. In the mass spectrum shown in Fig. 4, several y and b fragment ions are annotated. These ions have a wide range of charge states between $4^+$ and $16^+$. As it can be seen on the Supplementary Figure S3, these different fragments can overlap, where the mass difference between the isotopes of the two different fragments is 20 mDa. Thus, a mass spectrometer with high mass resolving power (resolving power of ~47 000 at mass 936) is needed to resolve these overlapping charge states and to properly deconvolute the spectrum. Also, these overlapping peaks show the advantage of using the selected isotope images which makes it possible to image the individual charge states with the added benefit of the full image contrast from the summed individual isotope peak intensities. Since this was an MS/MS imaging experiment, the images of the fragment ions can be plotted. The examples on Fig. 4 show the selected isotope images of the ions at m/z 885.7678 ($y_{56}^{7+}$ fragment) and at m/z 926.5439 ($b_{123}^{14+}$ fragment). Although these fragments show the same distribution on the tissue, the possibility to map the distribution of protein fragments from a top-down proteomics experiment on a tissue section offers the prospect to image the distribution of different proteoforms. This can give new insight in the mechanism of biological processes where protein modifications are involved.

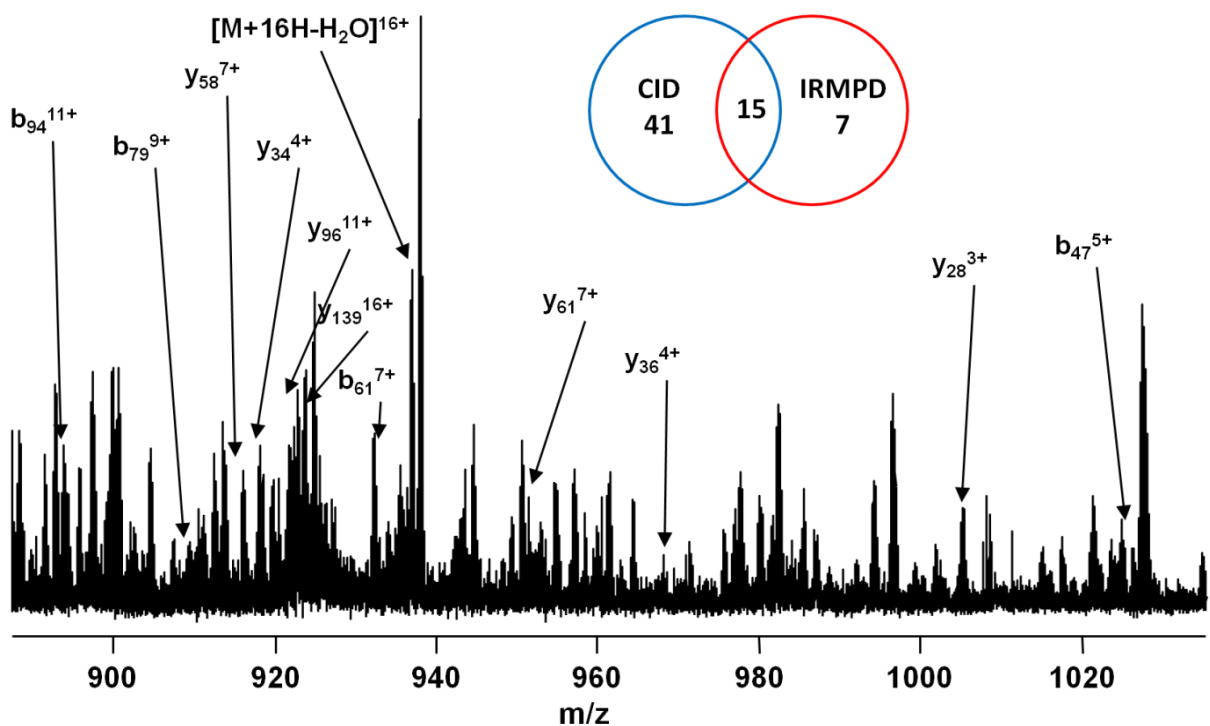

**Figure 5**

Figure 5 presents the results of an IRMPD MS/MS imaging experiment of the same precursor ion at m/z 883 (the broadband spectrum is shown in the supplementary material). After deconvolution of the summed spectrum, the database search resulted in the identification of the protein as hemoglobin α, with a p value of $1.25*10^{-5}$ in absolute mass search mode with 12 annotated fragments and $2*10^{-11}$ in single protein search with 19 identified fragments of the acetylated hemoglobin α. This result is in agreement with the result of the CID fragmentation. Thus, it improves the confidence of the protein identification. IRMPD has a similar fragmentation mechanism as CID. Thus, similar b and y ions were expected, as shown in Supplementary Table S1 and S2, which list the annotated fragments from the CID and IRMPD experiments, respectively. There is a substantial overlap between the fragment ions produced by the two fragmentation methods. Nevertheless, seven fragments are exclusively present in the IRMPD spectrum; see the Venn diagram in Fig. 5. Thus, the two fragmentation methods provide complementary datasets. However, the fragmentation efficiency of IRMPD is lower than of CID as it is proven by the lower number of fragments produced by IRMPD fragmentation.

**Conclusion**

This work presents the first example of top-down mass spectrometry imaging with a LAESI ion source. The protein identified in this work is hemoglobin which is among the most abundant proteins in a tissue sample. For the analysis of lower abundance proteins further instrumental developments are needed. The most important of these is to increase the sensitivity of the LAESI FT-ICR system. This can includes the improvement of the ion source and capacitive coupling of the FT-ICR cell which is estimated to result in two-fold sensitivity increase. Further possible improvements also include different commonly used tissue washing methods to remove lipids to enhance protein and peptide signal. Also, the investigation of potential IR matrices, such as glycerol or succinic acid, might result in further increases in the sensitivity of the LAESI FT-ICR system. These improvements are also required to be able to decrease the laser spot size and thus to increase the spatial resolution of the system. In addition, ECD/ETD fragmentation of intact proteins would be a good compliment to CID and IRMPD fragmentation. It has a different fragmentation mechanism compared to CID or IRMPD and it produces c and z protein fragments and is more gentle to allow labile PTMs to be retained. Thus it would provide complementary information to the other fragmentation methods.

This paper presents that imaging and identification of intact proteins and their modifications is achievable directly from tissue with the combination of high mass resolution mass spectrometers and an ambient imaging ion source. Multiply charged proteins were fragmented with IRMPD and ECD and identified directly from liquid standards. In addition, multiply charged proteins directly from frozen tissue sections were imaged by LAESI FT-ICR MS and identified without any additional sample preparation. In addition, a post-translational modification (acetylation) was identified for the first time directly from tissue and the position of the post-translational modification in the protein sequence was determined. This MS-based top-down proteomics imaging approach opens up new possibilities in biological research. The study of the distribution of protein proteoforms and labile post translational modifications directly from tissue provide new insight in the role of the different proteoforms in biological processes and diseases.

**Acknowledgements**

This work is part of the research program of the Foundation for Fundamental Research on Matter (FOM), which is part of the Netherlands Organisation for Scientific Research (NWO). This publication was supported by the Dutch national program COMMIT and the Netherlands Proteomics Center. The authors are thankful to Marco Konijnenburg and Ivo Klinkert for their support with the data processing, Julia Jungmann and Marco Seynen for their help with the experimental setup and Mike Senko for help with the LTQ-FT hardware.

**Figure Legends**

Figure 1 Mass spectra from the IRMPD (a) and ECD (b) fragmentation of Cytochrome C measured with a LAESI FT-ICR MS directly from liquid droplets.

Figure 2 Summed full mass spectrum from LAESI FT-ICR MS imaging of a mouse lung section with the charge state series of a 15 kDa protein (*) and 15.6 kDa protein (•) marked (a) and two multiply charged protein ions between m/z 820 and 830 (b). The inset at the top-left shows the resolved isotope structure of the protein ion at m/z 883

Figure 3 Optical (g), selected ion images (a, c, e) and selected isotope images (b, d, f) from a LAESI FT-ICR MS imaging experiment of a mouse lung section. The MS images show the distribution of the ions at m/z 822 (15.6 kDa: a, b), 827 (5 kDa: c, d) and 883 (15.6 kDa: e, f)

Figure 4 Zoomed summed mass spectrum from the CID MS imaging experiment of a mouse lung tissue section. Insets show selected isotope images of two fragments and the optical image of the lung section

Figure 5 Zoomed summed mass spectrum from an IRMPD MS imaging experiment of a mouse lung section. The Venn-diagram shows the number of unique fragments annotated from the CID and the IRMPD experiment

**Supplementary material**

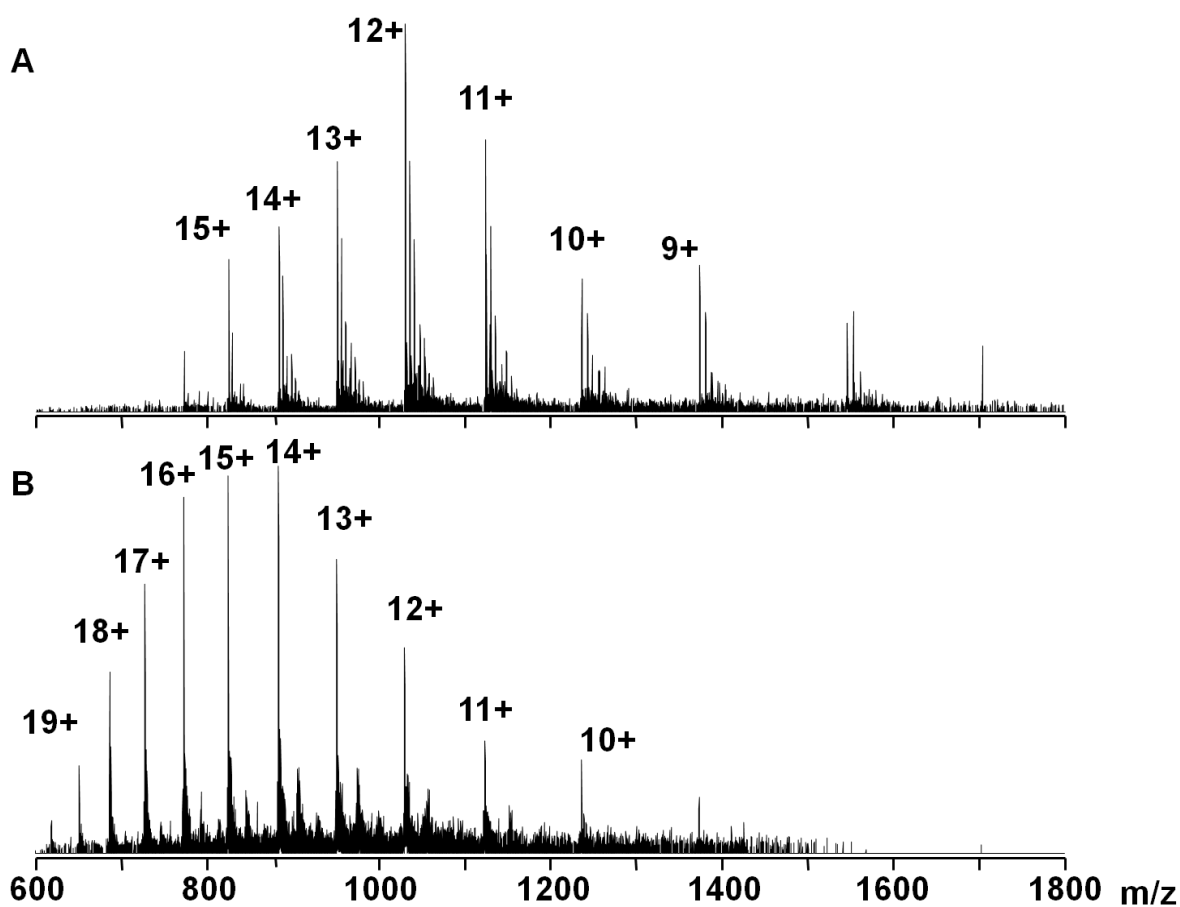

Supplementary figure S1 Comparison of the mass spectrum of Cytochrome C standard solution acquired with ESI (a) and LAESI (b). The electrospray spectra were measured with the IonMax source at a flow rate of 5 ul/min and an electrospray voltage of 4.2 kV.

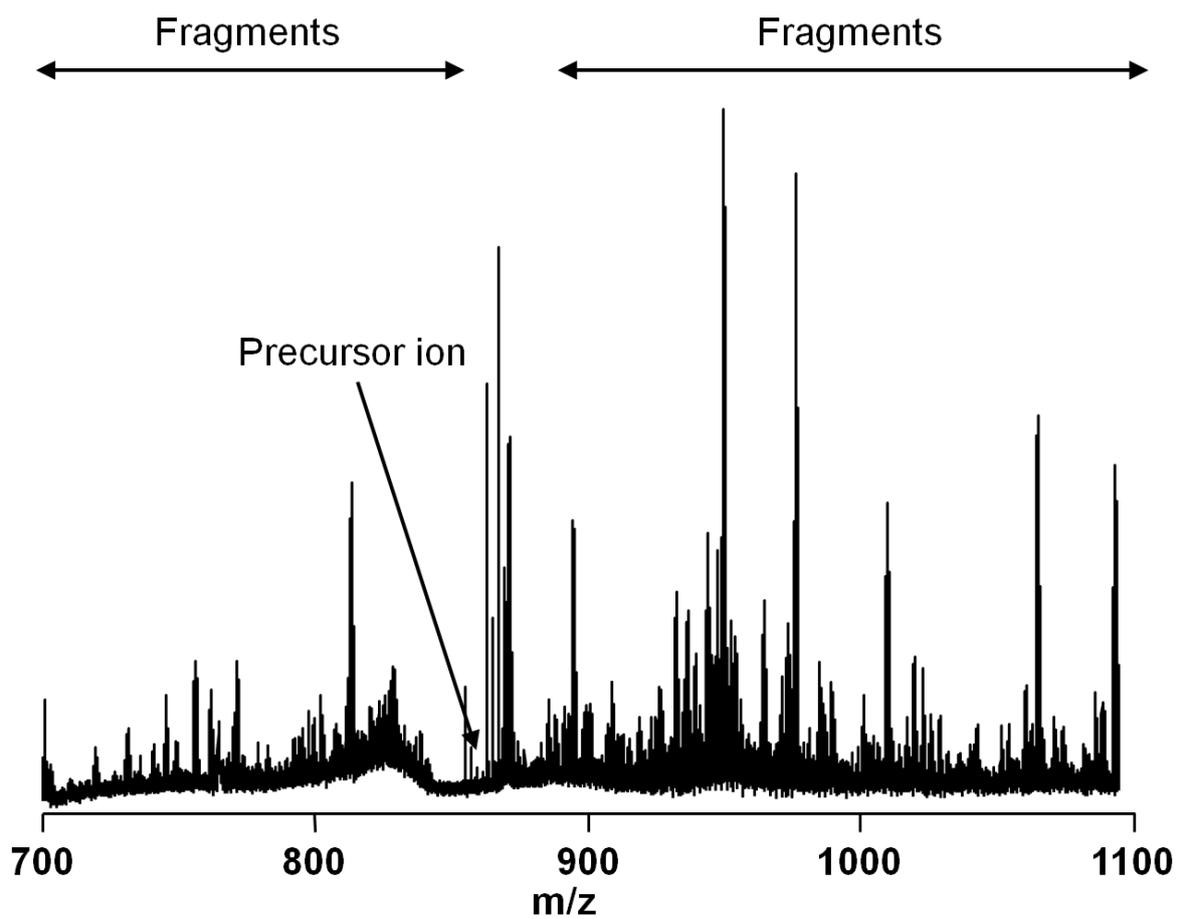

Supplementary figure S2 Full summed mass spectrum from the CID MS imaging experiment of mouse lung.

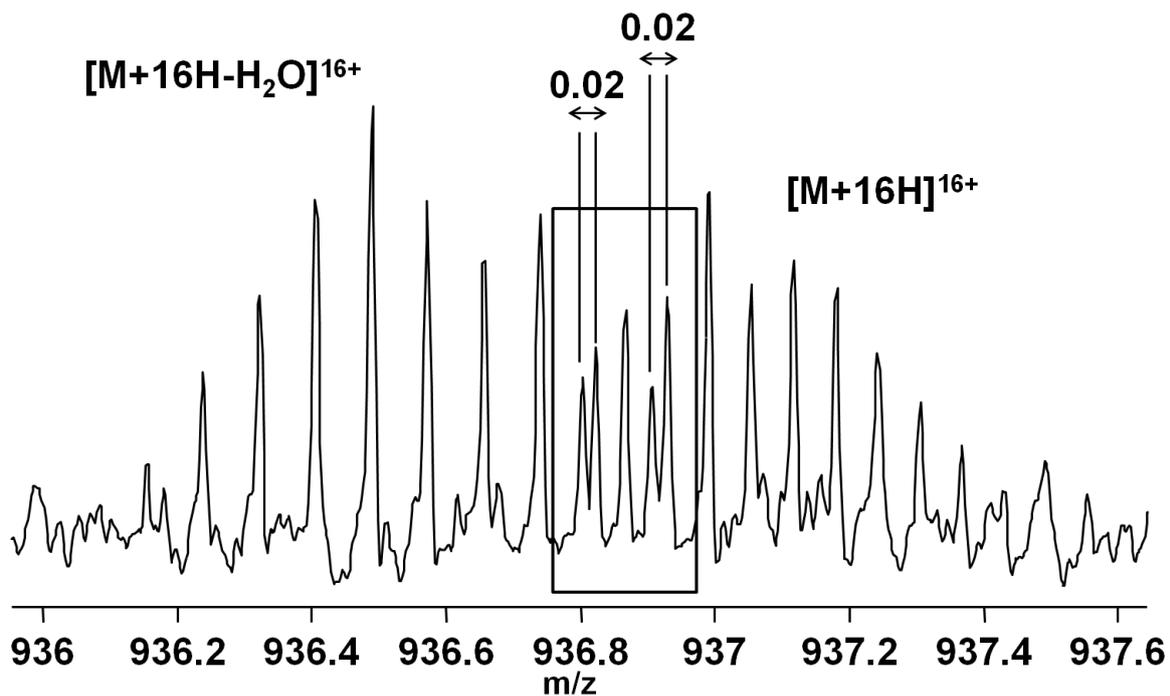

Supplementary figure S3 Zoom mass spectrum from the CID MS imaging experiment of mouse lung showing two overlapping fragment ions.

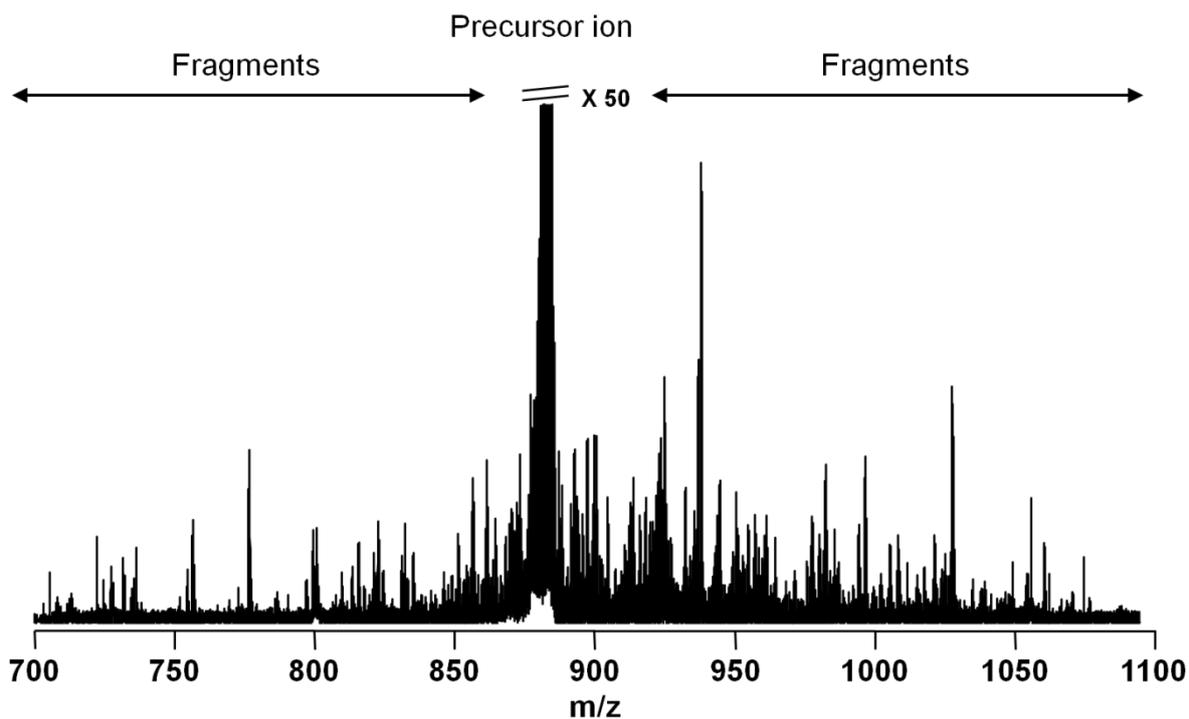

Supplementary figure S3 Full summed spectrum from the IRMPD MS imaging experiment of mouse lung.

**Supplementary Table S1 List of annotated CID fragments**

| Annotated ions | Charge state | Experimental monoisotopic mass | Deconvoluted experimental mass | Theoretical monoisotopic mass | Error (Da) | Error (ppm) |
|---|---|---|---|---|---|---|
| b45 | 6 | 808.3980 | 4850.388 | 4850.39 | -0.0040 | -0.82 |
| b47 | 6 | 852.0772 | 5112.463 | 5112.49 | -0.0239 | -4.68 |
| b60 | 7 | 912.1633 | 6385.143 | 6385.16 | -0.0129 | -2.01 |
| b61 | 8 | 814.1519 | 6513.215 | 6513.25 | -0.0357 | -5.47 |
| b63 | 8 | 835.2934 | 6682.347 | 6683.36 | 0.0414 | 6.20 |
| b64 | 8 | 849.6713 | 6797.371 | 6798.38 | 0.0380 | 5.59 |
| b74 | 8 | 968.3539 | 7746.831 | 7746.85 | -0.0190 | -2.45 |
| b78 | 8 | 1015.998 | 8127.980 | 8129.04 | -0.0045 | -0.55 |
| b79 | 9 | 911.0048 | 8199.044 | 8200.07 | 0.0220 | 2.68 |
| b103 | 12 | 907.5473 | 10890.57 | 10891.6 | 0.0482 | 4.42 |
| b116 | 13 | 942.1721 | 12248.24 | 12249.3 | 0.0377 | 3.07 |
| b118-$H_2O$ | 13 | 959.8743 | 12478.37 | 12479.4 | 0.0602 | 4.82 |
| b118 | 14 | 892.5979 | 12496.37 | 12497.4 | 0.0543 | 4.34 |
| b118 | 13 | 961.2596 | 12496.37 | 12497.4 | 0.0586 | 4.69 |
| b123 | 14 | 926.5439 | 12971.61 | 12972.6 | 0.0436 | 3.36 |
| b126 | 14 | 949.0561 | 13286.79 | 13287.8 | 0.0717 | 5.40 |
| b127 | 14 | 958.2056 | 13414.88 | 13415.9 | 0.0703 | 5.24 |
| b130 | 15 | 916.4662 | 13746.99 | 13747.0 | -0.0558 | -4.06 |
| y12 | 2 | 655.3613 | 1310.723 | 1310.72 | 0.0013 | 0.96 |
| y19 | 3 | 695.0475 | 2085.142 | 2085.15 | -0.0065 | -3.10 |
| y20 | 3 | 740.7350 | 2222.205 | 2222.21 | -0.0028 | -1.25 |
| y23 | 3 | 829.7870 | 2489.361 | 2489.37 | -0.0051 | -2.03 |
| y28 | 4 | 755.1494 | 3020.598 | 3020.60 | -0.0015 | -0.50 |
| y28 | 3 | 1006.867 | 3020.601 | 3020.60 | 0.0018 | 0.59 |
| y29 | 4 | 789.4123 | 3157.649 | 3157.66 | -0.0088 | -2.79 |
| y32 | 4 | 862.9439 | 3451.776 | 3452.79 | 0.0402 | 11.6 |
| y34 | 5 | 733.1817 | 3665.908 | 3666.92 | 0.0412 | 11.2 |
| y34 | 4 | 916.7271 | 3666.908 | 3666.92 | -0.0094 | -2.57 |
| y35 | 4 | 941.2432 | 3764.973 | 3765.99 | 0.0375 | 9.95 |
| y35 | 5 | 753.1947 | 3765.973 | 3765.99 | -0.0127 | -3.37 |
| y36 | 5 | 775.6125 | 3878.062 | 3879.07 | 0.0428 | 11.0 |
| y36 | 4 | 969.7637 | 3879.055 | 3879.07 | -0.0155 | -4.01 |
| y38 | 5 | 819.0312 | 4095.156 | 4095.16 | -0.0075 | -1.83 |
| y47 | 6 | 854.9565 | 5129.739 | 5130.75 | 0.0393 | 7.67 |
| y47 | 5 | 1025.948 | 5129.739 | 5130.75 | 0.0400 | 7.79 |
| y56 | 8 | 775.0436 | 6200.349 | 6200.36 | -0.0155 | -2.50 |
| y56 | 7 | 885.7678 | 6200.374 | 6200.36 | 0.0098 | 1.58 |

| Annotated ions | Charge state | Experimental monoisotopic mass | Deconvoluted experimental mass | Theoretical monoisotopic mass | Error (Da) | Error (ppm) |
|---|---|---|---|---|---|---|
| y57 | 7 | 902.1964 | 6315.375 | 6315.39 | -0.0166 | -2.62 |
| y58 | 7 | 914.6278 | 6402.395 | 6402.42 | -0.0289 | -4.51 |
| y58 | 6 | 1067.068 | 6402.410 | 6402.42 | -0.0138 | -2.15 |
| y58 | 8 | 800.3015 | 6402.412 | 6402.42 | -0.0115 | -1.79 |
| y59 | 7 | 930.6409 | 6514.486 | 6515.51 | 0.0292 | 4.48 |
| y59 | 8 | 814.3112 | 6514.489 | 6515.51 | 0.0324 | 4.97 |
| y60 | 8 | 823.1901 | 6585.521 | 6586.54 | 0.0266 | 4.04 |
| y61 | 7 | 953.2221 | 6672.555 | 6673.58 | 0.0288 | 4.32 |
| y62 | 7 | 969.3717 | 6785.602 | 6786.66 | -0.0081 | -1.19 |
| y63 | 7 | 979.5217 | 6856.652 | 6857.70 | 0.0049 | 0.72 |
| y63 | 8 | 857.2091 | 6857.673 | 6857.70 | -0.0249 | -3.64 |
| y74-$H_2O$ | 8 | 989.2410 | 7913.928 | 7915.18 | -0.2004 | -25.3 |
| y74 | 8 | 991.4943 | 7931.954 | 7933.19 | -0.1868 | -23.5 |
| y74 | 8 | 991.5241 | 7932.193 | 7933.19 | 0.0522 | 6.58 |
| y96 | 11 | 921.3953 | 10135.35 | 10136.3 | 0.0572 | 5.64 |
| y128 | 15 | 911.6003 | 13674.00 | 13674.0 | -0.0298 | -2.18 |
| y139 | 16 | 923.3462 | 14773.54 | 14774.6 | 0.0092 | 0.62 |
| M-$H_2O$ | 16 | 935.4185 | 14966.70 | 14967.6 | 0.1189 | 7.94 |
| precursor ion | 16 | 936.5423 | 14984.68 | 14985.6 | 0.0889 | 5.94 |

Supplementary Table S2 List of annotated IRMPD fragments

| Annotated ions | Charge state | Experimental monoisotopic mass | Deconvoluted experimental mass | Theoretical monoisotopic mass | Error (Da) | Error (ppm) |
|---|---|---|---|---|---|---|
| b94 | 11 | 895.9072 | 9854.979 | 9855.98 | 0.0470 | 4.77 |
| y23 | 3 | 829.7873 | 2489.362 | 2489.37 | -0.0043 | -1.72 |
| y28 | 3 | 1006.864 | 3020.593 | 3020.60 | -0.0063 | -2.10 |
| y28 | 4 | 755.1479 | 3020.591 | 3020.60 | -0.0076 | -2.51 |
| y32 | 4 | 862.9419 | 3451.768 | 3452.79 | 0.0323 | 9.34 |
| y34 | 4 | 916.7274 | 3666.910 | 3666.92 | -0.0082 | -2.23 |
| y35 | 5 | 753.1932 | 3765.966 | 3765.99 | -0.0202 | -5.37 |
| y36 | 4 | 969.5154 | 3878.062 | 3879.07 | 0.0422 | 10.9 |
| y47 | 5 | 1026.147 | 5130.733 | 5130.75 | -0.0169 | -3.29 |
| y47 | 6 | 855.1212 | 5130.727 | 5130.75 | -0.0228 | -4.43 |
| y56 | 7 | 885.7657 | 6200.360 | 6200.36 | -0.0048 | -0.78 |
| y56 | 8 | 775.0420 | 6200.336 | 6200.36 | -0.0283 | -4.56 |
| y58 | 7 | 914.6281 | 6402.397 | 6402.42 | -0.0267 | -4.17 |
| y59 | 7 | 930.6412 | 6514.488 | 6515.51 | 0.0314 | 4.83 |
| y61 | 7 | 953.3633 | 6673.543 | 6673.57 | -0.0312 | -4.68 |
| y61 | 8 | 834.0672 | 6672.537 | 6673.57 | 0.0132 | 1.97 |
| y81 | 10 | 860.0537 | 8600.537 | 8601.58 | 0.0108 | 1.26 |
| y96 | 11 | 921.3956 | 10135.35 | 10136.3 | 0.0607 | 5.99 |
| y125 | 15 | 886.7871 | 13301.81 | 13302.8 | 0.0193 | 1.45 |
| M-2H$_2$O | 17 | 879.3894 | 14949.62 | 14950.7 | -0.0382 | -2.55 |
| M-H$_2$O | 17 | 880.4510 | 14967.67 | 14968.7 | -0.0015 | -0.10 |
| precursor ion | 17 | 881.5081 | 14985.64 | 14986.7 | -0.0417 | -2.78 |